\documentclass[twocolumn]{aastex631}
\usepackage{CJKutf8}
\usepackage{amssymb}
\usepackage{amsmath}
\usepackage{multirow}
\usepackage{subfigure}
\usepackage{hyperref}
\usepackage[nameinlink,noabbrev]{cleveref}
\usepackage[normalem]{ulem}
\usepackage{xcolor}

\newcommand\redsout{\bgroup\markoverwith{\textcolor{red}{\rule[0.5ex]{2pt}{0.4pt}}}\ULon}

\newcommand{\asiaa}{Institute of Astronomy and Astrophysics, Academia
  Sinica, No.1, Sec. 4, Roosevelt Rd, Taipei 10617, Taiwan}

\newcommand{\cfa}{Center for Astrophysics | Harvard \& Smithsonian, 60
  Garden Street, Cambridge, MA 02138, USA}

\newcommand{\upenn}{Department of Physics and Astronomy, University of
  Pennsylvania, 209 South 33rd Street, Philadelphia, PA 19125, USA}

\newcommand{\nrc}{Herzberg Astronomy and Astrophysics Research Centre,
  National Research Council of Canada, 5071 West Saanich Road,
  Victoria, BC V9E 2E7, Canada}

\newcommand{\uvic}{Department of Physics and Astronomy, University of
  Victoria, Victoria, BC V8W 2Y2, Canada}

\newcommand{\kobe}{Department of Planetology, Kobe University, Kobe
  657-8501, Japan}

\newcommand{\ueoh}{University of Occupational and Environmental Health
  1-1 Iseigaoka, Yahata, Kitakyusyu 807-8555, Japan}

\newcommand{\chiba}{Planetary Exploration Research Center, Chiba
  Institute of Technology 2-17-1 Tsudanuma,Narashino, Chiba 275-0016,
  Japan}

\newcommand{\pmo}{CAS Key Laboratory of Planetary Sciences, Purple
  Mountain Observatory, Chinese Academy of Sciences, Nanjing 210023,
  China}

\newcommand{\cas}{CAS Center for Excellence in Comparative
  Planetology, CAS, Hefei 230026, China}

\newcommand{\subaru}{Subaru Telescope, National Astronomical
  Observatory of Japan, 650 North A`ohoku Place, Hilo, HI 96720, USA}

\newcommand{\kasi}{Korea Astronomy and Space Science Institute, 776
  Daedeok-daero, Yuseong-gu, Daejeon 34055, Republic of Korea}

\newcommand{\jsa}{Japan Spaceguard Association, Bisei Spaceguard
  Center 1716-3 Okura, Bisei, Ibara, Okayama 714-1411, Japan}

\newcommand{\ccanaoj}{Center for Computational Astrophysics, National
  Astronomical Observatory of Japan, Osawa 2-21-1, Mitaka, Tokyo,
  181-8588, Japan}

\newcommand{\campion}{Campion College and the Department of Physics,
  University of Regina, 3737 Wascana Parkway, Regina, SK S4S 0A2,
  Canada}

\newcommand{\nanjing}{School of Astronomy and Space Science, Nanjing
  University, 163 Xianlin Avenue, Nanjing 210023, China}

\newcommand{\keylab}{Key Laboratory of Modern Astronomy and
  Astrophysics in Ministry of Education, Nanjing University, Nanjing
  210023, China}

\newcommand{\kindai}{School of Interdisciplinary Social and Human
  Sciences, Kindai University, Shinkamikosaka 228-3, Higashiosaka-shi,
  Osaka, 577-0813, Japan}

\newcommand{\ncu}{Institute of Astronomy, National Central University,
  No. 300, Zhongda Rd., Zhongli Dist., Taoyuan City 32001, Taiwan}

\newcommand{\shao}{Shanghai Astronomical Observatory, Chinese Academy
  of Sciences, 80 Nandan Road, Shanghai 200030, China}

\newcommand{\iucaa}{Inter-University Centre for Astronomy and
  Astrophysics, Ganeshkhind, Pune 411007, India}

\newcommand{\ipmu}{Kavli Institute for the Physics and Mathematics of
  the Universe (WPI), 5-1-5 Kashiwanoha 2778583, Japan}

\submitjournal{PSJ}

\shorttitle{FOSSIL JT Spin Rate Limit}
\shortauthors{Chang et al.}

\begin{document}

\title{FOSSIL: I. The Spin Rate Limit of Jupiter Trojans}

\correspondingauthor{Chan-Kao Chang}
\email{rckchang@asiaa.sinica.edu.tw}


\author[0000-0003-1656-4540]{Chan-Kao Chang
  (\begin{CJK*}{UTF8}{bkai}章展誥\end{CJK*})}
\affiliation{\asiaa}

\author[0000-0001-7244-6069]{Ying-Tung Chen
  (\begin{CJK*}{UTF8}{bkai}陳英同\end{CJK*})}
\affiliation{\asiaa}

\author[0000-0001-6680-6558]{Wesley C. Fraser}
\affiliation{\nrc}
\affiliation{\asiaa}

\author[0000-0002-3286-911X]{Fumi Yoshida
  (\begin{CJK*}{UTF8}{min}吉田二美\end{CJK*})}
\affiliation{\ueoh} \affiliation{\chiba}

\author[0000-0003-4077-0985]{Matthew J. Lehner}
\affiliation{\asiaa}
\affiliation{\upenn}
\affiliation{\cfa}

\author[0000-0001-6491-1901]{Shiang-Yu Wang
  (\begin{CJK*}{UTF8}{bkai}王祥宇\end{CJK*})}
\affiliation{\asiaa}

\author[0000-0001-7032-5255]{JJ Kavelaars}
\affiliation{\nrc}
\affiliation{\uvic}

\author[0000-0003-4797-5262]{Rosemary E. Pike}
\affiliation{\cfa}

\author[0000-0003-4143-8589]{Mike Alexandersen}
\affiliation{\cfa}

\author[0000-0002-0549-9002]{Takashi Ito
  (\begin{CJK*}{UTF8}{min}伊藤孝士\end{CJK*})}
\affiliation{\ccanaoj}

\author{Young-Jun Choi
  (\begin{CJK*}{UTF8}{mj}최영준\end{CJK*})}
\affiliation{\kasi}

\author[0000-0001-8214-5147]{A. Paula {Granados Contreras}}
\affiliation{\asiaa}

\author[0000-0003-3435-7596]{Youngmin JeongAhn
  (\begin{CJK*}{UTF8}{mj}정안영민\end{CJK*})}
\affiliation{\kasi}

\author{Jianghui Ji
  (\begin{CJK*}{UTF8}{gbsn}季江徽\end{CJK*})}
\affiliation{\pmo}

\author[0000-0002-4787-6769]{Myung-Jin Kim
  (\begin{CJK*}{UTF8}{mj}김명진\end{CJK*})}
\affiliation{\kasi}

\author[0000-0001-5368-386X]{Samantha M. Lawler}
\affiliation{\campion}

\author{Jian Li
  (\begin{CJK*}{UTF8}{gbsn}黎健\end{CJK*})}
\affiliation{\nanjing}
\affiliation{\keylab}

\author[0000-0003-3827-8991]{Zhong-Yi Lin
  (\begin{CJK*}{UTF8}{bkai}林忠義\end{CJK*})}
\affiliation{\ncu}

\author[0000-0003-0926-2448]{Patryk Sofia Lykawka}
\affiliation{\kindai}

\author{Hong-Kyu Moon
  (\begin{CJK*}{UTF8}{mj}문홍규\end{CJK*})}
\affiliation{\kasi}

\author{Surhud More}
\affiliation{\iucaa}
\affiliation{\ipmu}

\author[0000-0002-0792-4332]{Marco Mu\~{n}oz-Guti\'{e}rrez}
\affiliation{\asiaa}

\author[0000-0002-4383-8247]{Keiji Ohtsuki
  (\begin{CJK*}{UTF8}{min}大槻圭史\end{CJK*})}
\affiliation{\kobe}

\author[0000-0003-4143-4246]{Tsuyoshi Terai}
\affiliation{\subaru}

\author[0000-0001-7501-8983]{Seitaro Urakawa
  (\begin{CJK*}{UTF8}{min}浦川聖太郎\end{CJK*})}
\affiliation{\jsa}

\author{Hui Zhang}
\affiliation{\shao}

\author{Hai-Bin Zhao
  (\begin{CJK*}{UTF8}{gbsn}赵海斌\end{CJK*})}
\affiliation{\pmo}
\affiliation{\cas}

\author{Ji-Lin Zhou
    (\begin{CJK*}{UTF8}{gbsn}周济林\end{CJK*})}
\affiliation{\nanjing}

\collaboration{28}{The FOSSIL Collaboration}

\begin{abstract}
Rotation periods of 53~small (diameters $2~\mathrm{km} < D < 40$~km)
Jupiter Trojans (JTs) were derived using the high-cadence light curves
obtained by the FOSSIL phase I survey, a Subaru/Hyper Suprime-Cam
intensive program. These are the first reported periods measured for
JTs with $D < 10$~km. We found a lower limit of the rotation period
near 4~hr, instead of the previously published result of 5~hr
\citep{Ryan2017,Szabo2017,Szabo2020} found for larger JTs. Assuming a
rubble-pile structure for JTs, a bulk density of $\approx$0.9~g\;cm$^{-3}$
is required to withstand this spin rate limit, consistent with the
value \replaced{1.08 \citep{Mueller2010}}{$\sim0.8 - 1.0$~g\;cm$^{-3}$ \citep{Marchis2006Natur.439..565M, Mueller2010,Buie2015AJ....149..113B, Berthier2020Icar..35213990B}} derived from the binary JT system, (617)
Patroclus–Menoetius system.
\end{abstract}

\keywords{minor planets, asteroids: general}

\section{Introduction} \label{sct:intro}

\begin{figure*}[t]
  \centering
  \includegraphics[width=.8\linewidth]{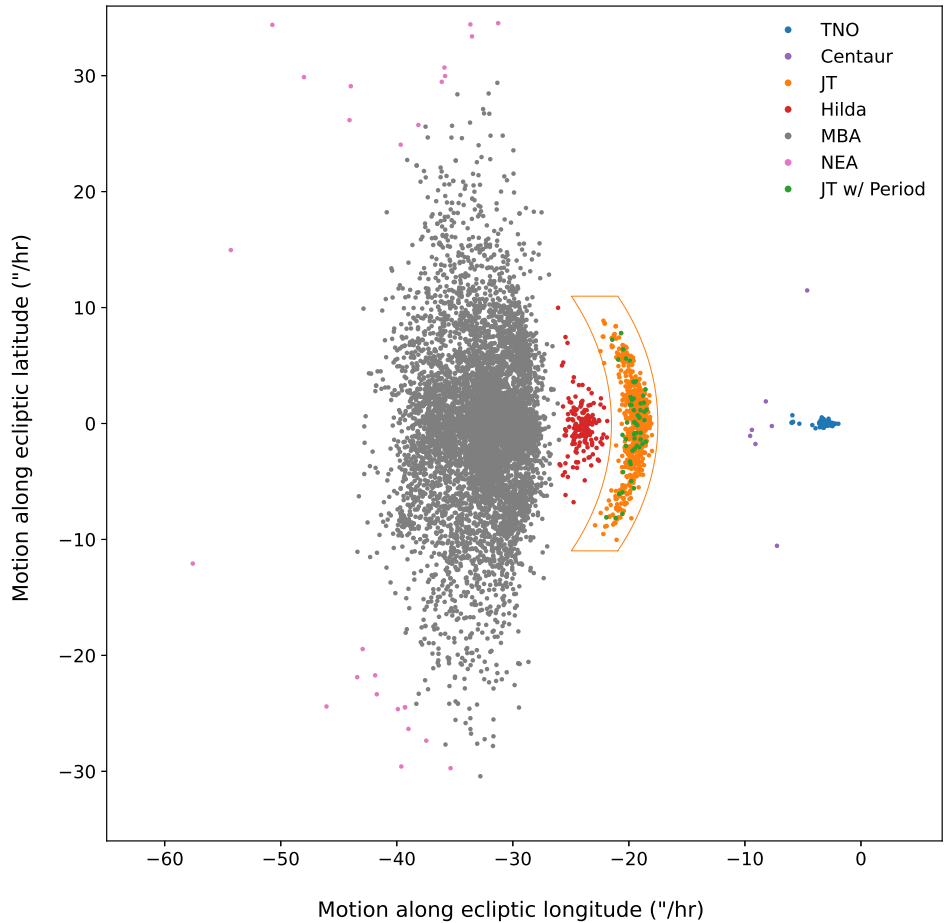}
  \caption{Motions along the ecliptic longitude and latitude of the
    moving objects detected in the FOSSIL observations. The moving
    objects are classified as outlined in \citet{Yoshida2017}. The JTs
    discussed in this paper are indicated by orange dots, and the JTs
    with fit rotation periods are indicated in green. \added{Note that the area
    outlined by orange are used to select JTs from FOSSIL survey,
    where two ellipses and an upper bound in absolute value of 11
    arcsec/hr in motion along ecliptic latitude are used.
    The ellipses can be described as $x_i, y_i = u_i+29*\cos(\theta) , 23.2*\sin(\theta)$,
    where $u$ are -46.5 and -50.5 arcsend/hr for $i=1$ and 2, respectively.}}
  \label{fig:mvplot}
\end{figure*}

The FOSSIL\footnote{\url{https://www.fossil-survey.org}} Survey
(Formation of the Outer Solar System: An Icy Legacy) is an intensive
survey program using Subaru/Hyper Suprime-cam (HSC). The goal of the
program is to measure the populations and characteristics of Jupiter
Trojans (JTs) and the various dynamical sub-populations of the small
bodies in the Trans-Neptunian region. The results of this survey
program will provide important clues to our understanding of the
formation and evolution of our Solar System. A major scientific goal
of the initial phase of the survey is to obtain high-cadence
lightcurves of small JTs and measure their rotation periods.


\begin{deluxetable*}{lccccccccccc}
  \tabletypesize{\scriptsize}
  \setlength{\tabcolsep}{0.1in}
  \tablecaption{Details of the observations for each survey block. \label{tab:obs}}
  \tablewidth{0pt}
  \tablehead{ \colhead{Block} & \colhead{RA} & \colhead{Dec} & \colhead{JT} & \colhead{Number of} & \colhead{Filter} & \colhead{Exposure} & \colhead{Limiting} & \colhead{Cadence} & \colhead{Date} & \colhead{Exposures} & \colhead{Time} \vspace{-.3cm}\\
  \colhead{ID} & \colhead{(deg)} & \colhead{(deg)} & \colhead{Cloud} & \colhead{Pointings} & \colhead{} & \colhead{Time (s)} & \colhead{Magnitude} & \colhead{(min)} & \colhead{} & \colhead{per Pointing} & \colhead{Span (hr)}}
\startdata
19Apr &  197.526 &  -6.763 & L5 & 5 &  $g$  &   90 & 24.5 & 10 & 2019-04-10 & 53 & 8.8 \\
20May &  224.351 & -14.596 & L5 & 2 &  $r2$ &  300 & 25.6 & 11 & 2020-05-19 & 23 & 3.8 \\
      &          &         &    &   &       &      & 25.7 &    & 2020-05-20 & 21 & 3.8 \\
20Aug &  341.656 &  -6.039 & L4 & 3 &  $r2$ &  300 & 25.6 & 16 & 2020-08-21 & 16 & 4.5 \\
      &          &         &    &   &       &      & 25.6 &    & 2020-08-22 & 15 & 4.0 \\
      &          &         &    &   &       &      & 25.5 &    & 2020-08-23 & 15 & 4.1 \\
20Oct &   10.119 &   5.754 & L4 & 3 &  $r2$ &  150 & 25.5 & 15 & 2020-10-14 & 24 & 3.4 \\
      &         &          &    &   &       &      & 25.4 &    & 2020-10-15 & 24 & 3.4 \\
      &         &          &    &   &       &      & 24.0 &    & 2020-10-16 &  8 & 1.0 \\
      &         &          &    &   &       &      & 24.0 &    & 2020-10-17 &  8 & 1.0 \\
\enddata
\end{deluxetable*}

JTs are a population of asteroids co-orbiting with Jupiter near its
L4~and L5~Lagrangian points. Because the orbits of JTs are relatively
stable, it is believed that their properties hold important \replaced{primitive
information from the early stages of the formation of our solar
system}{information about the primitive Solar System}. JTs could
have been formed at their present locations during
the formation of Jupiter \citep{Marzari1998,Fleming2000}, or they
formed somewhere else during the early stages of the solar system
formation and were then captured into their current locations as
Trojans during migration of the giant planets
\citep{Fernandez1984,Malhotra1995,Morbidelli2005,Lykawka2010,Nesvorny2013}.
Comparative studies of the overall physical properties between JTs and
other small body populations are crucial to our understanding of the
origin of JTs and the formation of our solar system.

A significant amount of previous work has been completed in order to
better understand the JT population. Two major spectral groups, i.e.,
red (D-type) and less red (P-type), have been identified within this
population \citep{Emery2011, Wong2014, Wong2015} \added{and only a
  very small fraction of C-types were also found
  \citep{Demeo2013Icar..226..723D}}. However, it is not clear how this
color bi-modality is related to other small body populations in the
solar system which also have dichotomous colors. Although the size
distribution of JTs is different from that of the Main Belt Asteroids
(MBAs) \citep{Yoshida2017}, this difference could be a result of
either different primordial origins or different evolutionary
histories. Finally, several measurements of the JT binary fraction
have been reported \citep{Mann2007, Sonnett2015, Ryan2017, Szabo2017,
  Nesvorny2020}, but this value is very uncertain, with estimates in
the range of 10\% to 30\%

The bulk densities and interior structures of JTs will also provide
useful insight when compared with other small body populations. In
addition to probing these properties for individual objects through
space missions or binary searches, overall estimations can be made
through their common spin-rate limit which can be identified from a
rotation period survey. Thanks to the availability of wide-field
cameras, this application has been extensively used on MBAs in the
past few years. It is believed \citep{Chapman1978NASCP2053..145C,
  Davis1985Icar...62...30D, Weissman1986Natur.320..242W} that MBAs
with diameters $1~\mathrm{km} \lesssim D \lesssim 100$~km are
gravitational aggregates (rubble-pile structures). These asteroids can
thus be destroyed if they spin too fast, and consequently have an
upper limit for their spin rates. \citet{Harris1996} first reported a
2~hr rotation period lower limit for MBAs of diameters $D > 150$~m and
suggested that these asteroids have rubble-pile structures with a
lower limit on their bulk densities of $\sim$3~g$\;$cm$^{-3}$. This
2~hr rotation period limit has consistently been seen in more recent
data sets \citep{Masiero2009, Chang2015, Chang2016,
  Chang2019}. Interestingly, more than two dozen super fast rotators
(SFRs), asteroids with $D > 300$~m and rotation periods $<2$~hr, have
been found \citep[see][and references therein]{Chang2019}. Unless
these objects have extremely high bulk densities, rubble-pile
structures could not survive such high rotation rates, indicating
cohesive force is required in addition to gravity to preserve the
structures of these objects \citep{Holsapple2007, Hirabayashi2015,
  Hu2021}.

While the wide-field surveys for asteroid rotation periods referenced
above have helped to understand their bulk densities and interior
structures, this kind of survey has not been conducted for
JTs. \citet{Ryan2017}, \citet{Szabo2017,Szabo2020} reported a possible
rotation period lower limit of $\approx$5~hr for JTs using the K2 data
set. However, their JT samples were limited to diameters $D > 10$~km,
and \replaced{the averaged spin rate of JTs may become faster around
  $D \approx 10$~km as is seen among MBAs
  \citep{Warner2009}. Therefore, it is necessary to extend the sample
  to smaller JTs ($D < 10$ km) to see if this 5~hr rotation period
  limit remains.}{these relatively large JTs have probably not been
  accelerated by the Yarkovsky–OKeefe–Radzievskii–Paddack (YORP)
  effect \citep{Rubincam2000} to reach their spin-rate limit.}

\added{The YORP effect is a mechanism to change the spin state of an
  object due to sunlight absorption and re-emission. Assuming circular
  orbits, the acceleration of the YORP effect on the spin rate can be
  expressed
\begin{equation}
  \frac{d\omega}{dt} \propto \frac{1}{\rho a^2 D^2},
\end{equation}
  where $\rho$ is the bulk density, $a$ is the orbital semimajor axis,
  and $D$ is the diameter of that moving object
  \citep{Rozitis2013MNRAS.430.1376R}. The YORP time-scale to double
  the spin rate of a 10~km diameter MBA is around a few hundred
  million years, and subsequently they have been sufficiently
  influenced by the YORP effect to reach their 2~hr spin-rate
  limit. In comparison, for JTs with similar densities and diameters
  of $D \sim 10$~km, it would take about a billion years to reach such
  a high angular velocity, and therefore the previously measured JT
  spin rate could very well be underestimated. Spin-rate measurements
  of smaller JTs are thus necessary to obtain a more accurate estimate
  of the true spin-rate limit.}

To achieve this goal we used Subaru and HSC \citep{Miyazaki2018,
  Komiyama2018, Kawanomoto2018, Furusawa2018} to conduct a wide-field
survey, from which dense lightcurves with durations from 1 to 3~nights
were collected to measure rotation periods for small JTs ($D \leq 10$
km). A total of 53~rotation periods \added{out of 1241 observed JTs}
were obtained by this survey.

This article is organized as follows. The observations, data
reduction, and lightcurve extraction are described in
\autoref{sct:obs}. The rotation period analysis is discussed in
\autoref{sct:period-analysis}. The results and discussion are
presented in \autoref{sct:results}, and a summary is given in
\autoref{sct:summary}.

\begin{deluxetable}{lcc}
  \tabletypesize{\scriptsize}
  \setlength{\tabcolsep}{0.1in}
  \tablecaption{Photometric Measurements for JT data presented in this
    paper.
    \label{tab:lightcurve}}
  \tablewidth{0pt}
  \tablehead{ \colhead{JD} & \colhead{Mag} & \colhead{Mag Error}}
  \startdata
\multicolumn{3}{c}{FASP03010029} \\
\multicolumn{3}{c}{(156294) 2001 WU66} \\
2458583.857932 &  20.9790 &  0.0062 \\
2458583.864924 &  20.9308 &  0.0061 \\
2458583.885930 &  20.8650 &  0.0059 \\
2458583.892958 &  20.8545 &  0.0058 \\
2458583.900007 &  20.8459 &  0.0058 \\
2458583.907036 &  20.8728 &  0.0059 \\
2458583.914078 &  20.9097 &  0.0059 \\
2458583.921106 &  20.9317 &  0.0060 \\
2458583.928126 &  21.0006 &  0.0062 \\
2458583.938686 &  21.0513 &  0.0065 \\
2458583.945697 &  21.1222 &  0.0069 \\
2458583.952704 &  21.2509 &  0.0070 \\
2458583.959705 &  21.3866 &  0.0078 \\
2458583.966701 &  21.4131 &  0.0078 \\
2458583.973695 &  21.4045 &  0.0078 \\
2458583.980718 &  21.3482 &  0.0080 \\
2458583.987715 &  21.3897 &  0.0076 \\
2458583.994707 &  21.3218 &  0.0074 \\
2458584.001710 &  21.2038 &  0.0072 \\
2458584.008693 &  21.0953 &  0.0070 \\
2458584.015689 &  21.0485 &  0.0068 \\
2458584.022678 &  20.9955 &  0.0068 \\
2458584.029678 &  20.9469 &  0.0066 \\
2458584.036667 &  20.9121 &  0.0065 \\
2458584.043656 &  20.8892 &  0.0063 \\
2458584.050641 &  20.8427 &  0.0062 \\
2458584.057636 &  20.8129 &  0.0062 \\
2458584.071627 &  20.8376 &  0.0063 \\
2458584.078620 &  20.8561 &  0.0065 \\
2458584.085612 &  20.8340 &  0.0067 \\
2458584.092621 &  20.9180 &  0.0069 \\
2458584.099606 &  20.9786 &  0.0070 \\
2458584.106602 &  20.9651 &  0.0078 \\
2458584.113589 &  21.0144 &  0.0079 \\
\enddata
\tablecomments{This is an example for a single object; all
  measurements from all 53~objects are available in machine-readable
  format.}
\end{deluxetable}

\section{Observations and Data Reduction} \label{sct:obs}
High-cadence observations were performed on four blocks of pointings
targeting the L4 and L5 JT clouds. The details of observations can be
found in \autoref{tab:obs}. Observations were conducted using
Subaru/HSC during 2019 April 10 (19Apr), 2020 May 18--19 (20May), 2020
August 20--22 (20Aug), and 2020 October 13--16 (20Oct). The $g$-band
filter was used for the 19Apr observations, and the $r2$-band was used
for the other three blocks. The observation time spans were roughly
8~hr for 19Apr and 4~hr each night for the observations conducted in
2020 (other than the last two nights of the 20Oct block, where the
time was reduced due to poor weather).

The 19Apr data is from a previous observing run, and was not part of
the original FOSSIL proposal. However, given that everyone of the
proposers for the 19Apr observations is a member of the FOSSIL
collaboration, these data were combined with the FOSSIL data set.

FOSSIL was originally awarded four nights in both May (2020A) and
September (2020B) when the L5 and L4 JT clouds, respectively, were at
opposition. However, we lost three of our four nights of scheduled
observations in 2020A because of the shutdown of Maunakea due to the
COVID-19 pandemic, and our 2020B nights were rescheduled to August and
October due to necessary Subaru maintenance which had been deferred
due to the pandemic. The change in our observing schedule necessitated
the cancellation of our planned JT color measurements, but we were
still able to make useful JT lightcurve measurements during the time
we managed to observe.

The number of selected pointings and the exposure time of each frame
were adjusted for each block as we learned from our experience from
the analysis of the previously observed blocks. For each block,
exposures were cycled through the selected pointings repeatedly
throughout each night. The typical limiting magnitudes were
24.5~mag for 19Apr and 25.4--25.7~mag for the others
(except for the last two nights of the 20Oct block observations where
the limiting magnitude dropped to 24~mag due to poor weather
conditions). A total of 13~pointings were used, for a total sky
coverage of 37.7~deg$^2$. Of this, 17.4~deg$^2$ covered the L4 cloud and
20.3~deg$^2$ covered L5.

All the images were processed using the official HSC pipeline,
hscPipe~v8.3 \citep{Bosch2018}, with astrometry and photometry
calibrated against the Pan-STARRS~1 catalog \citep{Chambers2017}. For
each pointing, hscPipe was used to build a template image in order to
produce differential images. The differential images were then
processed by the same pipeline to generate source catalogs of
potential moving objects.

Since the observations were carried out using relatively long exposure
times, the images of the moving objects with relatively short
geocentric distance were trailed. In order to improve the photometry
for the trailing moving objects, the trailed source fitting software
package TRIPPy \citep{Fraser2016,trippy} was used to measure the
magnitudes of the moving object candidates. For each CCD in the HSC
focal plane, TRIPPy creates a point spread function (PSF) model for
each exposure based on the PSFs from a subset of stars on the same
chip. This PSF model is then used along with the measured rate of
motion of the relevant JT to create a trailing aperture for the moving
object. The background is calculated as the median pixel value in the
differential image from a set of pixels separated from the trailed PSF
based on the full width half maximum (FWHM) of the model PSF. An
aperture correction based on the FWHM is then applied to the resulting
photometric measurement. \added{Photometric uncertainties are
estimated based on the signal-to-noise ratio.}  The intra-night
detections of the moving objects would appear as linear sequences with
correlated epochs. The Hough transform \citep{Hough1959, Duda1972}, an
algorithm for line detection in images, was thus utilized to correlate
the linear intra-night detections and find moving objects. This
procedure is described in detail in \citet{Chang2019}.

\begin{figure}
  \centering
  \includegraphics[width=\linewidth]{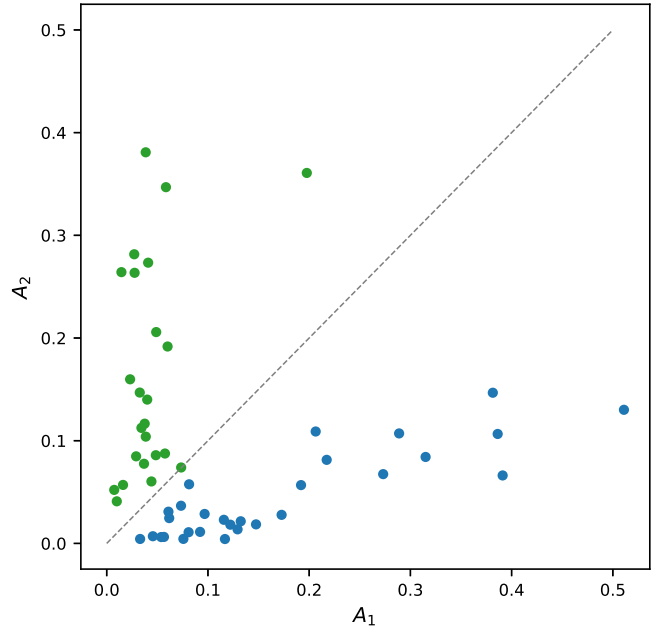}
  \caption{Amplitude $A_2$ vs $A_1$ for the folded lightcurves. Double
    peaked lightcurves (green dots) have $A_2 > A_1$ while the single
    peaked folded lightcurves (blue dots) have $A_2 < A_1$.}
  \label{fig:n1n2}
\end{figure}

\begin{figure}
  \includegraphics[width=\linewidth]{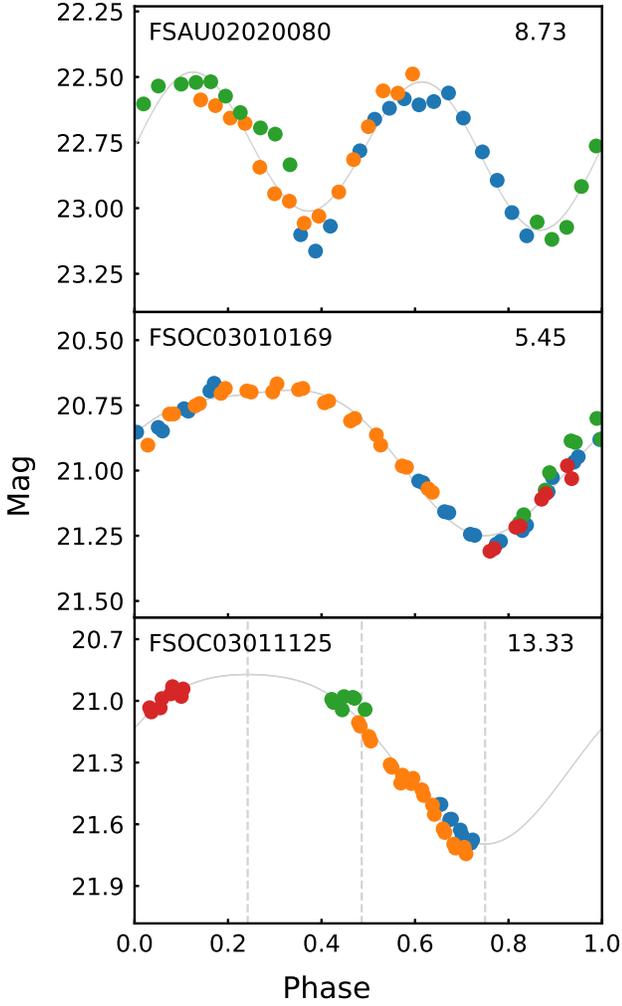}
  \caption{ The upper and mid panels show double peaked and single
    peaked folded lightcurves, respectively. The lower panel shows
    single peaked folded lightcurve with insufficient
    coverage. Dashed lines indicate the minimum, maximum, and points
    where the fit lightcurve crosses the mean magnitude
    $\bar{m}$. This lightcurve is rejected because there are no points
    in the fourth section. Note that the object ID and derived rotation
    period are indicated on each plot. Different colors represent data
    points obtained from different nights. The gray lines are the fitting
    results.}
    \label{fig:partial}
\end{figure}

Since observations were conducted near opposition, we are able to
use the rates of motion along the Ecliptic longitude and latitude
to distinguish different populations of moving objects
\citep[e.g.][]{Yoshida2017}. \added{The arc
lengths of the observed objects are limited to 1 to 3 days and,
therefore, cause a relatively large uncertainty in orbit
determination. Therefore, the Ecliptic motion is used to select JT
samples.} As shown in \autoref{fig:mvplot}, the observed moving
objects can be classified as MBAs, Hildas, JTs, and Trans-Neptunian
Objects (TNOs). Moreover, several Near Earth Asteroids (NEAs) and
Centaurs are also evident. We used the objects corresponding to the
orange points in \autoref{fig:mvplot} as our sample of JTs for
further \deleted{orbit determination and} rotation period analysis,
and the JTs for which we found periods are indicated by the green
dots. In total, 1241~JTs (hereinafter FOSSIL JTs) with detections
in five or more epochs were chosen, including 63 previously known
JTs. (Note that no rotation periods had been measured for these
63~objects.)

\begin{figure*}[t]
  \centering
  \includegraphics[width=0.8\linewidth]{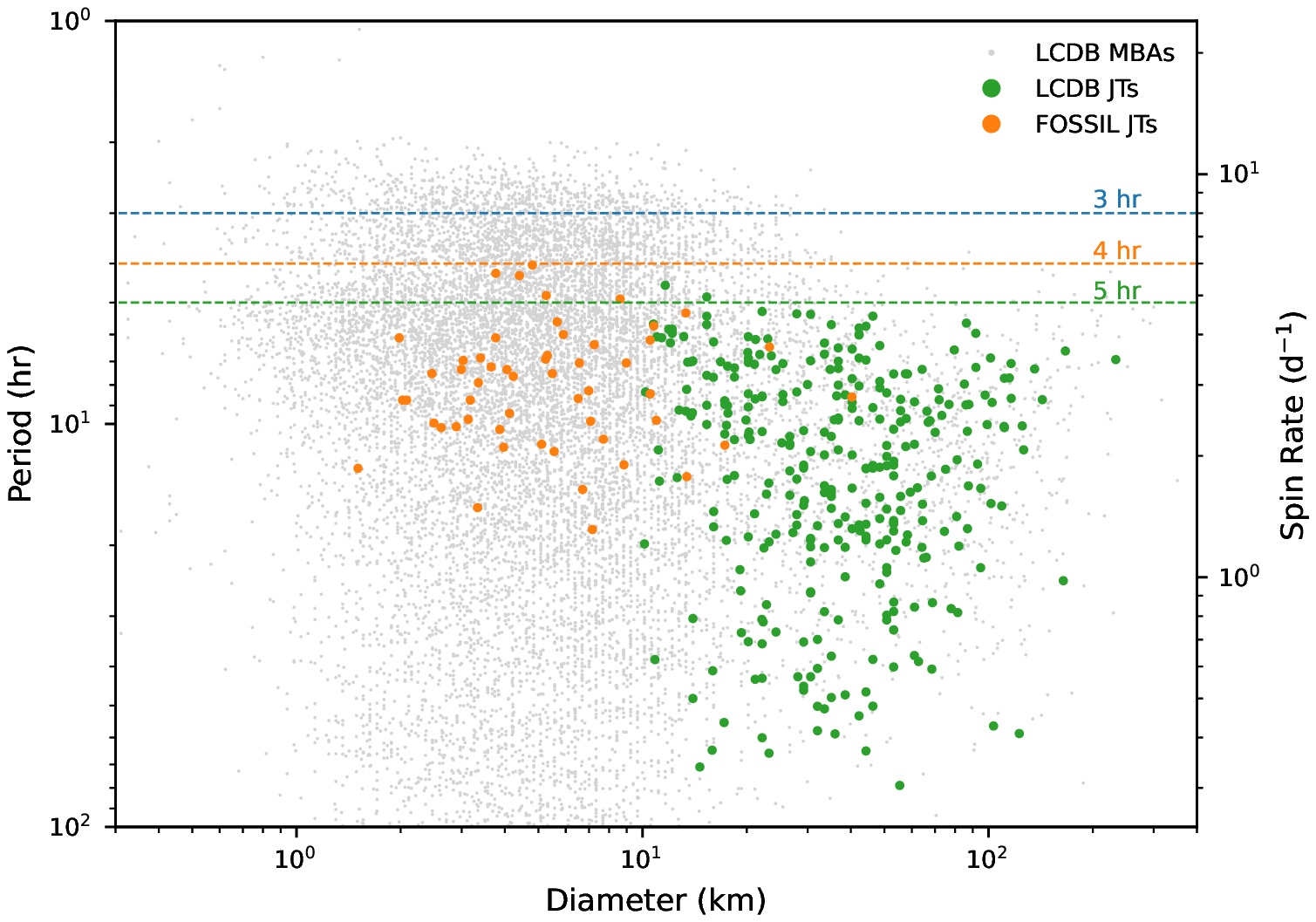}
  \caption{Rotation period and spin rate vs diameter for the FOSSIL
    JTs (orange dots). The same is shown for previously measured JTs
    \deleted{as well} (green dots) \added{and MBAs (background
      gray dots).}  The green, orange, and blue dashed lines indicate
    rotation periods of 5, 4, and 3 hr, respectively. Note that only
    the objects with full and half derived rotation periods in our
    samples are used in this plot. \deleted{All half periods are doubled for
    comparison to objects with full periods.} \added{Previously known
    JT and MBA rotation periods were obtained from the Asteroid Lightcurve
    Database (LCDB).}}
    \label{fig:dia-per}
\end{figure*}

In order to estimate the diameters of the FOSSIL JTs, the distance to
each object must be estimated. To that end, we assume a constant
semi-major axis of $a=5.2$~au and eccentricity $e = 0$ for each JT.
Since the phase angles of FOSSIL JTs only have small changes during
our observations, we simply estimate their absolute magnitudes using a
fixed $G$ slope of 0.15 in the $H$--$G$ system \citep{Bowell1989}.
Diameters were then estimated \citep{Yoshida2017} as

\begin{equation}\label{eq:dia-eq}
  \log D = 0.2m_\odot + \log2r - 0.5\log p - 0.2H,
\end{equation}

where $m_\odot$ is the apparent magnitude of the Sun, $r$ is the
heliocentric distance of Earth in the same unit as $D$, $p$ is the
geometric albedo, and $H$ is the absolute magnitude of the JT in the
observed band. We adopt $m_\odot = -27.04$ for the $r2$ band and
$-26.34$ for the $g$ band \citep{Willmer2018}, and we set $p=0.05$
\citep{Romanishin2018} for both bands.

\section{Rotation-Period Analysis}\label{sct:period-analysis}
To measure rotation period, we attempted to follow the method of
\citet{Harris1989} and performed a 2nd-order Fourier series fit to the
lightcurves of FOSSIL JTs\footnote{The correction for the
light-traveling time was not applied here because \deleted{it is}
\added{the resulting changes are} negligible for short time-span
surveys (i.e., 1 to 3 days).}:

\deleted{
  OLD EQ
}

\deleted{where $m_{i,j}$ are the apparent magnitudes in the observed
  band on night $i$, $t_j$ is the observing epoch for measurement $j$,
  $B_k$ and $C_k$ are the coefficients of the 2nd-order Fourier
  series, $P$ is the rotation period, $t_0$ is an arbitrary epoch, and
  $Z_i$ is an offset to account for the change in the phase angle over
  different nights of observation. The spin rate ($f = 1/P$) was
  explored from 0.25 to 50~d$^{-1}$ using a step size of
  0.01~d$^{-1}$. However, when performing these fits, we found that
  setting $Z_i$ to a free parameter often led to the fitting routine
  returning unphysical values of this offset in order to improve the
  quality of the fit. Given that all of our observations were made
  very close to opposition, we can ignore the phase angle effects and
  fit a mean magnitude which is the same for all nights of
  observation. The fit function thus becomes}

\begin{multline}\label{eq:FTeq2}
  m_j = \bar{m} + \sum_{k=1}^{2}\left\{ B_k\sin\left[\frac{2\pi k}{P}
    (t_j-t_0)\right] \right. \\
  \left.
  + C_k\cos\left[\frac{2\pi k}{P} (t_j-t_0)\right] \right\},
\end{multline}

\added{where $m_j$ are the apparent magnitudes in the
observed band, $t_j$ is the observing epoch for measurement $j$,
$B_k$ and $C_k$ are the coefficients of the 2nd-order Fourier
series, $P$ is the rotation period, $t_0$ is an arbitrary epoch,
and $\bar{m}$ is the mean magnitude of the JT. The spin rate ($f =
1/P$) was explored from 0.25 to 50~d$^{-1}$ using a step size of
0.01~d$^{-1}$.where $\bar{m}$ is the mean magnitude of the JT.}

To determine whether the algorithm gives a good fit to the lightcurve,
we calculate the difference between the reduced $\chi^2$ of the
best-fit period and that of a fit to the mean magnitude. We
found that when the difference is $>$2, the fitting shows a
convincing folded lightcurve.

Based on the assumption of ellipsoidal shapes for JTs, a folded
lightcurve with two minima and two maxima is expected. However, the
best-fit (i.e., the minimum reduced $\chi^2$) period of this algorithm
returns two types of folded lightcurves: double peaked and single
peaked. Two conditions can give rise to a single peaked lightcurve:
first, when all of the data are contained in the same half of the
phased double peaked lightcurve, and second, when the two halves of the
double peaked lightcurve are very similar. To distinguish between the
two cases, we look at the amplitudes of each phase $k$ of the Fourier
series fit

\begin{equation}
  A_k = (B_k^2 + C_k^2)^\frac{1}{2}.
\end{equation}

For a double peaked lightcurve, the amplitude of the $k=2$ Fourier
component is larger, with a smaller correction by the $k=1$
component. When a single peaked folded lightcurve is found, the
opposite is true. We can then thus distinguish between the two cases
by defining a folded lightcurve as double peaked when $A_2 > A_1$ and
single peaked when $A_1 \ge A_2$. \autoref{fig:n1n2} shows a plot of
$A_1$ vs $A_2$ for the JTs where a good fit was found, and
\autoref{fig:partial} shows example double and single peaked folded
lightcurves. Note that in most cases it is obvious if the fit
lightcurve is single or double peaked, but there are some marginal
cases when the amplitude is low, and this method also facilitates
automation of the analysis.

When the best-fit period gives a single peaked lightcurve, the next
best local minimum of the reduced $\chi^2$ vs $P$ curve with a longer
period is selected as the preferred solution. However, this does not
always work well when there is not sufficiently full coverage of the
single peaked folded lightcurve. To eliminate such cases, we divide
each single peak folded lightcurve into four sections bounded by the
minimum, maximum, and the two points where the fit lightcurve crosses
the mean magnitude $\bar{m}$ (see \autoref{fig:partial}). We require
that there be at least two points in each section, otherwise we reject
the lightcurve since we cannot be confident of the fit
period. Finally, when using the second local minimum for the period,
in some cases unrealistic fit parameters are returned (e.g. a
lightcurve amplitude of 80~mag). In such cases, we have found that the
fit period is always more than three times the fit period for the
original single peaked folded lightcurve. We thus reject any fits
where the new period is three times longer than that from the original
\added{single peaked} fit. \added{ Uncertainties in the periods are
  estimated following the process described in
  \citet{Polishook2012MNRAS.421.2094P}.}


\begin{figure*}[t]
  \centering
  \includegraphics[width=.8\linewidth]{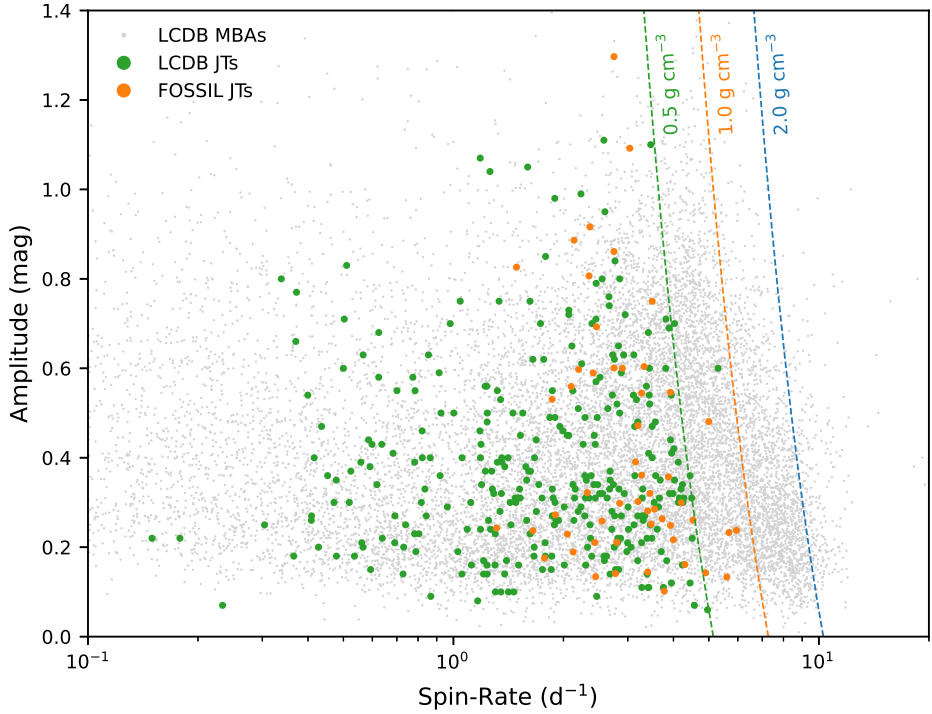}
  \caption{Spin rate vs amplitude of FOSSIL JTs (orange dots) and
    JTs  (green dots) and MBAs (gray dots) found in the LCDB.
    The green, orange, and blue dashed lines are the rotation
    period limits for rubble-pile asteroids with bulk densities
    of 0.5, 1, and 2~g\;cm$^{-3}$, respectively, calculated from
    \autoref{eq:plim}.}
  \label{fig:spin-amp}
\end{figure*}

From the lightcurves of the 1241~FOSSIL JTs, we obtained 40~double
peaked folded lightcurves which passed our selection criteria. In
addition, we found 13~single peaked folded lightcurves from which we
were able to recalculate double peaked lightcurves which passed the
cuts outlined above. The main reason for the low rate of successful
period fitting is due to the fact that most of the detections are of
fainter objects, and the lightcurves of these objects are therefore
too noisy to obtain an accurate fit given the short span of our
observations. In addition, due to the short time span of our
observations at each block, our analysis is insensitive to longer
period rotation curves.

Photometric data for these 53~lightcurves are presented in
\autoref{tab:lightcurve}. Diameters, rotation periods, lightcurve
amplitudes, and folded lightcurve fit parameters for each of these
objects are shown in \autoref{tab:jtpars} in the appendix. The folded
lightcurves of these objects are also shown in the appendix in
\Cref{fig:JTlcu2,fig:JTlchalf}.

\section{Results and Discussion} \label{sct:results}
\autoref{fig:dia-per} shows a plot of diameter vs rotation period for
the FOSSIL JTs where full and half rotation periods are found.  For
comparison, the values for previously measured \deleted{JT} rotation
periods for \added{JTs and MBAs} \footnote{Previously known
JT and MBA rotation periods were obtained from the Asteroid Lightcurve Database
\citep[LCDB][]{Warner2009} which can be found at
\url{http://www.minorplanet.info/lightcurvedatabase.html}.
\added{Note that only the rotation periods with quality code
  of $U \ge 2$ are used shown.}} are also shown. The FOSSIL data set
extends the range of diameters of JTs with measured rotation periods
from $D \gtrsim 10$~km down to $D \gtrsim 1$~km for the first time. We
note that there is a clear lack of long period detections in the
FOSSIL data. This is due to biases against long periods in our survey
arising from the short time span of observations at each block of
pointings.

In the sample of smaller diameter JTs found by FOSSIL, five of them
have rotation periods faster than the previously suggested 5-hr
limit, \deleted{with the shortest period being 4.03~hr.}\added{three
  out of which have rotation periods of $\sim 4$~hr. The diameters of
  these three 4-hr rotation period JTs are around 5 km, where the size
  range is expected to have sufficient YORP acceleration to reach the
  JT spin-rate limit, as mentioned in \autoref{sct:intro}. We also
  note that the upturn in the spin rates shown among MBAs with
  diameters around 30~to 40~km is possibly seen for JTs around diameters ~10
  to 20 km as well. This might indicate the diameter ranges where the
  YORP effect starts to affect the spin rates in both populations and,
  moreover, the diameter ranges follows simple relation of the YORP
  acceleration as discussd in \autoref{sct:intro}.}

Assuming a rubble-pile structure for JTs, the minimum bulk density to
withstand these spin rates can be calculated \citep{Harris1996}
using

\begin{equation}
  P = 3.3 \left(\frac{1 + \delta m}{\rho}\right)^\frac{1}{2},
  \label{eq:plim}
\end{equation}

where $P$ is the period in hr, $\rho$ is the bulk density in
g\;cm$^{-3}$, and $\delta m$ is the lightcurve amplitude in mag. We
estimate the lightcurve amplitude $\delta m$ as 95\% of the difference
between the brightest and fainted measurements for each object, given
that the folded lightcurve fits sometimes significantly overestimate
the amplitudes. \autoref{fig:spin-amp} shows a plot of spin rate vs
lightcurve amplitude for both the FOSSIL JTs and previously measured
JTs, along with limits on bulk density calculated from
\autoref{eq:plim}. Given the rotation rates measured for the FOSSIL
JTs, these objects need a bulk density of at least $\approx$0.9~g\;cm$^{-3}$, a
value consistent with the measurements of \replaced{1.08 \citep{Mueller2010}}{$\sim0.8 - 1.0$~g\;cm$^{-3}$ \citep{Marchis2006Natur.439..565M, Mueller2010,Buie2015AJ....149..113B, Berthier2020Icar..35213990B}} from the
binary JT system, (617)~Patroclus–Menoetius system
and much higher than that derived from the 5-hr spin-rate limit
\citep[i.e., $\sim$0.5~g\;cm$^{-3}$][]{Ryan2017, Szabo2017, Szabo2020,
Kalup2021ApJS..254....7K}.


\added{The favored formation scenario \citep{Nesvorny2013} suggests that JTs and dynamically excited Kuiper Belt Objects (KBOs) were populated from the same primordial planetesimal reservoir. Our findings point to a tension with this idea when considering the densities of KBOs;  small KBOs ($D\lesssim300$~km) have significantly lower bulk densities $\rho\sim0.7$~g\;cm$^{-3}$ \citep{Grundy2019Icar..334...30G} than Patroclus-Menoetius and JT bulk densities derived here assuming rubble pile structures. Possible solutions to this tension include the possibility that KBOs of the same size as the Trojans considered here have similarly higher densities. The fact that Patroclus-Menotius has a higher bulk density than similar sized KBOs however, disfavours this possibility. Collisional evolution \citep{Wong2014} may be responsible for raising the densities of KBOs \citep{Fraser2018AJ....156...23F}, but it remains to be seen whether collisional evolution sufficient to raise densities by $\sim\frac{1}{3}$ would not also unbind the Patroclus-Menotius system. }

\section{Summary and Conclusions} \label{sct:summary}
Using the Subaru/HSC, a wide-field high-cadence survey, part of which
was to measure rotation periods of small JTs, was conducted in 2019
and 2020. From this survey, we report the detection of 1241~JTs, only
63~of which are found in the MPC database. We were able to obtain
rotation periods for 53~of the 1241~JTs, the vast majority of which
were measured on objects with diameters $D<10$~km, an order of
magnitude smaller than previously accomplished. We found a number of
objects with periods near 4~hr, significantly lower than the suggested
limit of 5~hr. Under the assumption of a rubble-pile structure for
JTs, a bulk density of $\approx$0.9~g\;cm$^{-3}$ is required to maintain
their structure at that rotation period limit. This value is
comparable to the measurements of \replaced{1.08 \citep{Mueller2010}}{$\sim0.8 - 1.0$~g\;cm$^{-3}$ \citep{Marchis2006Natur.439..565M, Mueller2010,Buie2015AJ....149..113B, Berthier2020Icar..35213990B}} from the binary JT system, (617)
Patroclus–Menoetius.

\begin{acknowledgments}
This research is based on data collected at Subaru Telescope, which is
operated by the National Astronomical Observatory of Japan. We are
honored and grateful for the opportunity of observing the Universe
from Maunakea, which has the cultural, historical and natural
significance in Hawaii. Y.~JeongAhn acknowledges support from the
National Research Foundation of Korea(NRF) grant funded by the Korea
government(MSIT) (No. 2020R1C1C1012212). This work was supported in
part by JSPS KAKENHI grant Nos. JP18K13607 and JP21H00043. F. Yoshida acknowledges
support from MEXT/JSPS KAKENHI grant Nos. 20H04617 and 18K03730. Our
discussion for the FOSSIL survey project began at "Subaru Workshop on
Small Solar System Bodies" held at the Center for Planetary Science
(CPS) of Kobe University in November, 2018. We thank the Subaru
Telescope for its financial support for the workshop and the CPS,
especially Fumihiko Usui, for their warm hospitality and support.
\end{acknowledgments}


\begin{figure*}[t]
  \centering
  \includegraphics[width=\linewidth]{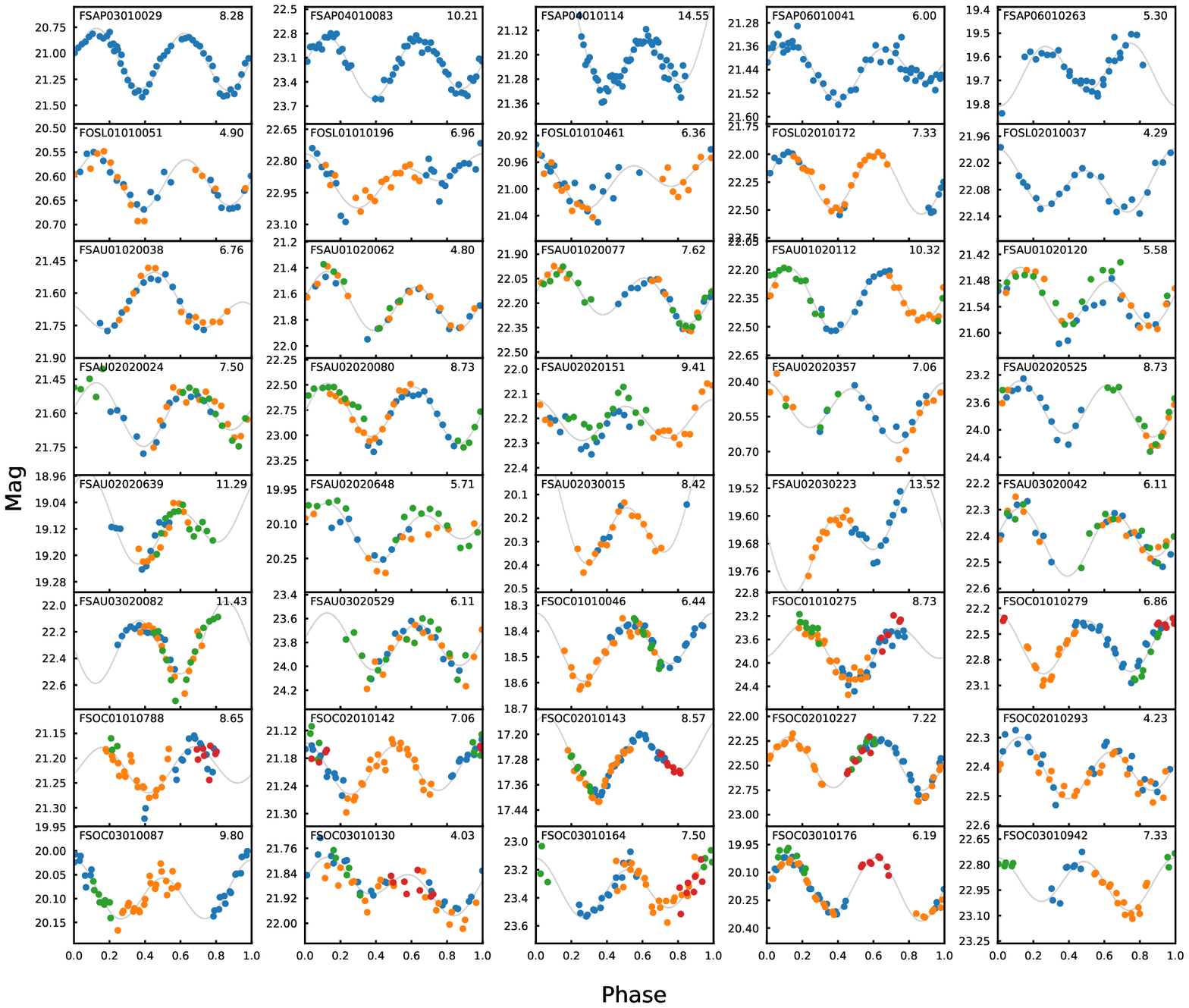}
  \caption{40~folded lightcurves of JTs where a double peak fit was
    returned. The object ID and derived rotation period are indicated
    on each plot. Different colors represent data points obtained from
    different nights. The gray lines are the fitting results. Photometric
    errors are too small to be seen in the plots.}
  \label{fig:JTlcu2}
\end{figure*}

\begin{figure*}[p]
  \centering
  \includegraphics[width=\linewidth]{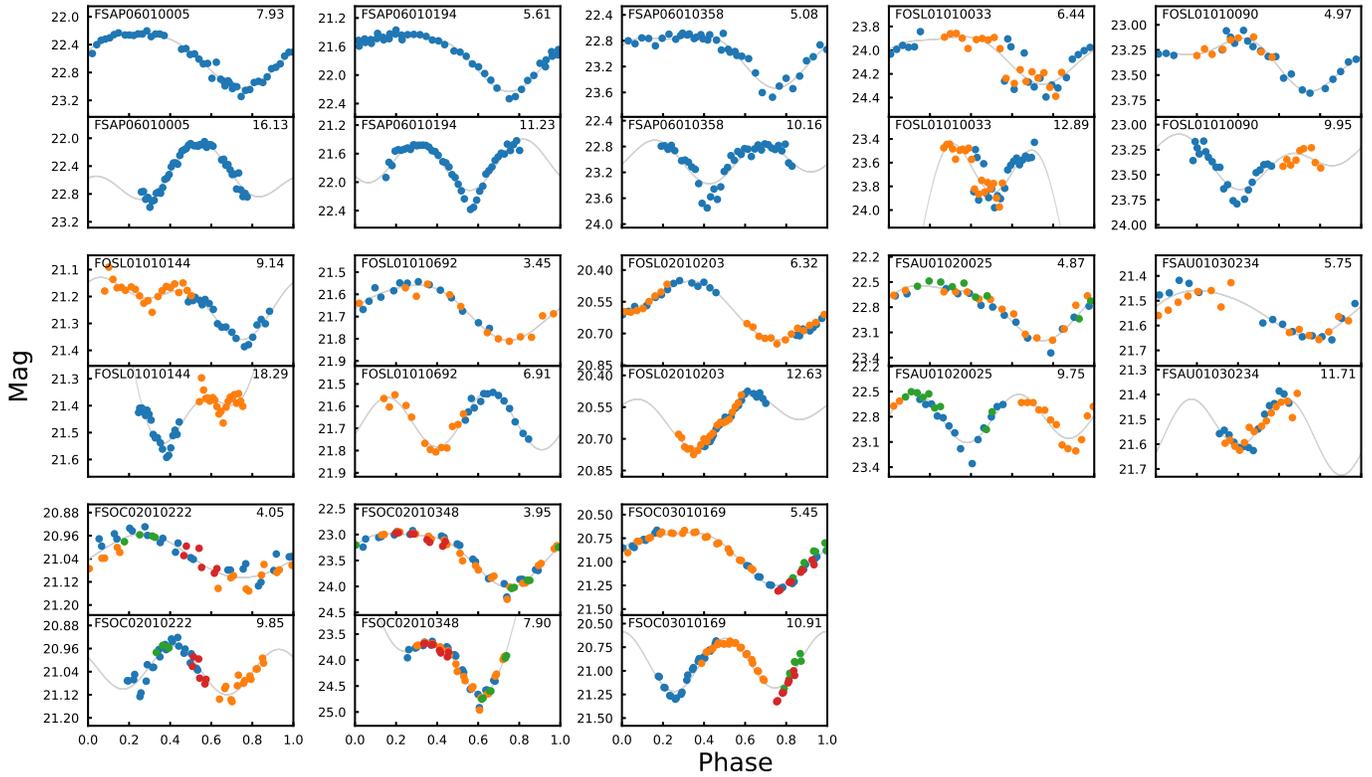}
  \caption{13~folded lightcurves of JTs where a single peak fit as
    returned. The top plots in each frame show the results from the
    original single peak fits, and the bottom plots show the folded
    lightcurves when a double peaked period was found in a second
    round of fitting. The object ID and half period are indicated on
    each plot. Symbols are the same as in
    \autoref{fig:JTlcu2}. Photometric errors are too small to be seen
    in the plots.}
  \label{fig:JTlchalf}
\end{figure*}

\clearpage

\begin{longrotatetable}
\startlongtable
\begin{deluxetable*}{lllDDDDDDDDDDD}
  \tabletypesize{\scriptsize}
  \setlength{\tabcolsep}{0.1cm}
  \tablecaption{Physical parameters for FOSSIL JTs for which rotation
    periods were obtained. \label{tab:jtpars}}
  \tablewidth{0pt}
  \tablehead{ \colhead{JT ID} & \colhead{MPC Designation} &
    \colhead{Block} & \multicolumn{2}{c}{$H$ (mag)} &
    \multicolumn{2}{c}{$D$ (km)} & \multicolumn{2}{c}{$\sigma_D$ (km)}
    & \multicolumn{2}{c}{$P$ (hr)} & \multicolumn{2}{c}{$\sigma_P$
      (hr)} & \multicolumn{2}{c}{$\delta m$ (mag) } &
    \multicolumn{2}{c}{$\bar{m}$ (mag)} & \multicolumn{2}{c}{$B_1$} &
    \multicolumn{2}{c}{$C_1$} & \multicolumn{2}{c}{$B_2$} &
    \multicolumn{2}{c}{$C_2$}}
  \decimals
  \startdata
FSAP03010029  &   (156294) 2001 WU66  &  19Apr  &   14.4  &    6.99 & 0.56  &   8.28  & 0.15   &  0.60 &  21.1 &   0.02 &  -0.02 &   0.12 &   0.25 \\
FSAP04010083  &                       &  19Apr  &   16.5  &    2.62 & 0.21  &  10.21  & 0.70   &  0.81 &  23.2 &  -0.05 &   0.03 &  -0.17 &  -0.30 \\
FSAP04010114  &                       &  19Apr  &   14.5  &    6.71 & 0.54  &  14.56  & 4.28   &  0.24 &  21.2 &  -0.03 &  -0.13 &  -0.09 &  -0.10 \\
FSAP06010041  &                       &  19Apr  &   14.7  &    5.90 & 0.48  &   6.00  & 0.23   &  0.22 &  21.4 &  -0.04 &  -0.00 &   0.05 &  -0.06 \\
FSAP06010263  &   (24022) 1999 RA144  &  19Apr  &   13.0  &   13.3  & 1.1   &   5.30  & 0.21   &  0.26 &  19.7 &   0.02 &  -0.03 &  -0.11 &  -0.03 \\
FOSL01010051  &    (523904) 1997 JF7  &  20May  &   13.9  &    8.62 & 0.70  &   4.898 & 0.049  &  0.14 &  20.6 &  -0.01 &  -0.00 &   0.02 &   0.05 \\
FOSL01010196  &                       &  20May  &   16.2  &    3.03 & 0.24  &   6.96  & 0.15   &  0.32 &  22.9 &  -0.04 &   0.06 &  -0.06 &  -0.05 \\
FOSL01010461  &            2011 PQ10  &  20May  &   14.3  &    7.25 & 0.59  &   6.36  & 0.13   &  0.10 &  21.0 &  -0.01 &   0.02 &  -0.03 &  -0.01 \\
FOSL02010172  &                       &  20May  &   15.6  &    4.06 & 0.33  &   7.328 & 0.056  &  0.54 &  22.2 &   0.02 &   0.02 &   0.25 &  -0.09 \\
FOSL02010037  &                       &  20May  &   15.4  &    4.41 & 0.36  &   4.28  & 0.35   &  0.13 &  22.1 &   0.00 &  -0.03 &   0.04 &  -0.04 \\
FSAU01020038  &                       &  20Aug  &   15.0  &    5.31 & 0.43  &   6.761 & 0.048  &  0.29 &  21.7 &   0.07 &  -0.00 &   0.00 &   0.09 \\
FSAU01020062  &                       &  20Aug  &   15.0  &    5.27 & 0.43  &   4.800 & 0.024  &  0.48 &  21.7 &  -0.01 &   0.06 &   0.03 &  -0.19 \\
FSAU01020077  &                       &  20Aug  &   15.5  &    4.23 & 0.34  &   7.619 & 0.061  &  0.39 &  22.2 &  -0.01 &  -0.04 &   0.11 &  -0.08 \\
FSAU01020112  &                       &  20Aug  &   15.7  &    3.87 & 0.31  &  10.32  & 0.11   &  0.32 &  22.4 &  -0.00 &   0.03 &  -0.13 &   0.07 \\
FSAU01020120  &                       &  20Aug  &   14.8  &    5.67 & 0.46  &   5.581 & 0.033  &  0.16 &  21.5 &  -0.00 &   0.02 &   0.04 &  -0.04 \\
FSAU02020024  &                       &  20Aug  &   14.9  &    5.48 & 0.44  &   7.500 & 0.059  &  0.30 &  21.6 &  -0.03 &  -0.02 &   0.11 &   0.02 \\
FSAU02020080  &                       &  20Aug  &   16.1  &    3.18 & 0.26  &   8.727 & 0.080  &  0.60 &  22.8 &  -0.01 &  -0.04 &   0.27 &   0.02 \\
FSAU02020151  &                       &  20Aug  &   15.5  &    4.13 & 0.33  &   9.41  & 0.19   &  0.26 &  22.2 &  -0.01 &   0.01 &   0.05 &  -0.05 \\
FSAU02020357  &            2015 CO51  &  20Aug  &   13.8  &    8.97 & 0.72  &   7.06  & 0.10   &  0.28 &  20.5 &   0.02 &  -0.03 &  -0.05 &   0.09 \\
FSAU02020525  &                       &  20Aug  &   17.0  &    2.08 & 0.17  &   8.727 & 0.080  &  0.86 &  23.7 &  -0.04 &   0.01 &  -0.30 &   0.23 \\
FSAU02020639  &   (100624) 1997 TR28  &  20Aug  &   12.4  &   17.3  & 1.4   &  11.29  & 0.13   &  0.19 &  19.1 &  -0.02 &  -0.07 &  -0.09 &   0.01 \\
FSAU02020648  &   (221909) 2008 QY14  &  20Aug  &   13.4  &   10.77 & 0.87  &   5.714 & 0.069  &  0.30 &  20.1 &  -0.06 &  -0.00 &   0.07 &  -0.05 \\
FSAU02030015  &  (286571) 2002 CR207  &  20Aug  &   13.5  &   10.52 & 0.85  &   8.42  & 0.15   &  0.30 &  20.2 &  -0.03 &  -0.15 &  -0.11 &  -0.13 \\
FSAU02030223  &   (257375) 2009 QZ47  &  20Aug  &   13.0  &   13.4  & 1.1   &  13.52  & 0.72   &  0.18 &  19.6 &  -0.09 &   0.02 &   0.09 &  -0.06 \\
FSAU03020042  &                       &  20Aug  &   15.7  &    3.76 & 0.30  &   6.115 & 0.079  &  0.25 &  22.4 &   0.02 &  -0.03 &  -0.10 &  -0.02 \\
FSAU03020082  &                       &  20Aug  &   15.6  &    3.97 & 0.32  &  11.43  & 0.28   &  0.56 &  22.3 &   0.08 &   0.01 &   0.17 &   0.18 \\
FSAU03020529  &                       &  20Aug  &   17.1  &    1.98 & 0.16  &   6.115 & 0.039  &  0.55 &  23.8 &  -0.05 &   0.01 &   0.18 &   0.10 \\
FSOC01010046  &                       &  20Oct  &   11.8  &   23.3  & 1.9   &   6.443 & 0.044  &  0.26 &  18.5 &   0.00 &   0.04 &  -0.08 &   0.07 \\
FSOC01010275  &                       &  20Oct  &   17.1  &    2.03 & 0.16  &   8.727 & 0.080  &  1.30 &  23.8 &  -0.16 &  -0.12 &   0.36 &   0.01 \\
FSOC01010279  &                       &  20Oct  &   15.9  &    3.40 & 0.27  &   6.857 & 0.049  &  0.75 &  22.6 &   0.04 &  -0.01 &   0.29 &   0.13 \\
FSOC01010788  &   (396413) 2014 ED23  &  20Oct  &   14.5  &    6.52 & 0.53  &   8.65  & 0.23   &  0.14 &  21.2 &   0.00 &   0.01 &   0.01 &   0.04 \\
FSOC02010142  &   (356253) 2009 UK77  &  20Oct  &   14.5  &    6.57 & 0.53  &   7.06  & 0.10   &  0.14 &  21.2 &   0.00 &   0.00 &   0.02 &   0.05 \\
FSOC02010143  &    (14690) 2000 AR25  &  20Oct  &   10.6  &   40.3  & 3.3   &   8.571 & 0.077  &  0.21 &  17.3 &  -0.01 &  -0.05 &  -0.07 &   0.04 \\
FSOC02010227  &                       &  20Oct  &   15.8  &    3.65 & 0.30  &   7.218 & 0.055  &  0.60 &  22.5 &  -0.01 &  -0.01 &   0.19 &  -0.18 \\
FSOC02010293  &                       &  20Oct  &   15.7  &    3.77 & 0.30  &   4.229 & 0.037  &  0.23 &  22.4 &   0.02 &   0.02 &  -0.01 &   0.08 \\
FSOC03010087  &                       &  20Oct  &   13.4  &   10.96 & 0.89  &   9.80  & 0.10   &  0.13 &  20.1 &   0.02 &   0.00 &  -0.01 &   0.05 \\
FSOC03010130  &                       &  20Oct  &   15.2  &    4.81 & 0.39  &   4.034 & 0.017  &  0.24 &  21.9 &  -0.03 &   0.03 &  -0.02 &   0.06 \\
FSOC03010164  &                       &  20Oct  &   16.6  &    2.46 & 0.20  &   7.500 & 0.059  &  0.47 &  23.3 &   0.04 &  -0.04 &   0.03 &  -0.17 \\
FSOC03010176  &   (263794) 2008 QQ37  &  20Oct  &   13.5  &   10.52 & 0.85  &   6.194 & 0.040  &  0.36 &  20.2 &  -0.01 &  -0.02 &   0.13 &   0.09 \\
FSOC03010942  &                       &  20Oct  &   16.2  &    2.99 & 0.24  &   7.33  & 0.16   &  0.36 &  22.9 &   0.01 &   0.03 &   0.13 &  -0.02 \\
FSAP06010005  &                       &  19Apr  &   16.0  &    3.34 & 0.27  &  16.13  & 1.81   &  0.83 &  22.7 &   0.21 &   0.14 &  -0.26 &   0.06 \\
FSAP06010194  &             2012 VQ5  &  19Apr  &   15.0  &    5.11 & 0.41  &  11.23  & 0.70   &  0.89 &  21.7 &   0.06 &  -0.00 &  -0.22 &  -0.23 \\
FSAP06010358  &                       &  19Apr  &   16.3  &    2.90 & 0.23  &  10.2   & 1.1    &  0.92 &  23.0 &  -0.06 &  -0.07 &   0.22 &   0.17 \\
FOSL01010033  &                       &  20May  &   17.7  &    1.51 & 0.12  &  12.89  & 0.54   &  0.53 &  24.4 &   0.52 &   0.39 &   0.43 &  -0.16 \\
FOSL01010090  &                       &  20May  &   16.6  &    2.50 & 0.20  &   9.95  & 0.30   &  0.59 &  23.3 &  -0.15 &  -0.05 &   0.14 &   0.07 \\
FOSL01010144  &                       &  20May  &   14.3  &    7.16 & 0.58  &  18.29  & 0.99   &  0.24 &  21.0 &   0.01 &   0.33 &  -0.08 &  -0.14 \\
FOSL01010692  &            2011 OR64  &  20May  &   15.0  &    5.25 & 0.42  &   6.91  & 0.15   &  0.25 &  21.7 &   0.00 &  -0.02 &  -0.08 &  -0.09 \\
FOSL02010203  &  (457150) 2008 FD133  &  20May  &   13.9  &    8.84 & 0.71  &  12.63  & 0.34   &  0.27 &  20.6 &  -0.03 &  -0.06 &   0.08 &   0.02 \\
FSAU01020025  &                       &  20Aug  &   16.1  &    3.14 & 0.25  &   9.75  & 0.19   &  0.69 &  22.8 &   0.01 &   0.03 &  -0.09 &   0.27 \\
FSAU01030234  &                       &  20Aug  &   14.9  &    5.55 & 0.45  &  11.71  & 0.60   &  0.23 &  21.6 &  -0.06 &  -0.02 &   0.07 &  -0.10 \\
FSOC02010222  &            2015 FP40  &  20Oct  &   14.3  &    7.08 & 0.57  &   9.84  & 0.10   &  0.21 &  21.0 &  -0.01 &   0.01 &  -0.02 &  -0.07 \\
FSOC02010348  &                       &  20Oct  &   16.0  &    3.35 & 0.27  &   7.90  & 0.13   &  1.09 &  22.7 &  -1.02 &  -0.43 &  -0.67 &  -0.04 \\
FSOC03010169  &  (295699) 2008 TC173  &  20Oct  &   14.2  &    7.71 & 0.62  &  10.91  & 0.13   &  0.60 &  20.9 &  -0.01 &  -0.04 &  -0.28 &  -0.09 \\
\enddata
\tablecomments{MPC designations are provided for previously known JTs.}
\end{deluxetable*}
\end{longrotatetable}

\bibliographystyle{aasjournal}
\bibliography{main}{}

\begin{thebibliography}{}
\expandafter\ifx\csname natexlab\endcsname\relax\def\natexlab#1{#1}\fi
\providecommand{\url}[1]{\href{#1}{#1}}
\providecommand{\dodoi}[1]{doi:~\href{http://doi.org/#1}{\nolinkurl{#1}}}
\providecommand{\doeprint}[1]{\href{http://ascl.net/#1}{\nolinkurl{http://ascl.net/#1}}}
\providecommand{\doarXiv}[1]{\href{https://arxiv.org/abs/#1}{\nolinkurl{https://arxiv.org/abs/#1}}}

\bibitem[{{Berthier} {et~al.}(2020){Berthier}, {Descamps}, {Vachier},
  {Normand}, {Maquet}, {Deleflie}, {Colas}, {Klotz}, {Teng-Chuen-Yu}, {Peyrot},
  {Braga-Ribas}, {Marchis}, {Leroy}, {Bouley}, {Dubos}, {Pollock}, {Pauwels},
  {Vingerhoets}, {Farrell}, {Sada}, {Reddy}, {Archer}, \&
  {Hamanowa}}]{Berthier2020Icar..35213990B}
{Berthier}, J., {Descamps}, P., {Vachier}, F., {et~al.} 2020, \icarus, 352,
  113990, \dodoi{10.1016/j.icarus.2020.113990}

\bibitem[{{Bosch} {et~al.}(2018){Bosch}, {Armstrong}, {Bickerton}, {Furusawa},
  {Ikeda}, {Koike}, {Lupton}, {Mineo}, {Price}, {Takata}, {Tanaka}, {Yasuda},
  {AlSayyad}, {Becker}, {Coulton}, {Coupon}, {Garmilla}, {Huang}, {Krughoff},
  {Lang}, {Leauthaud}, {Lim}, {Lust}, {MacArthur}, {Mandelbaum}, {Miyatake},
  {Miyazaki}, {Murata}, {More}, {Okura}, {Owen}, {Swinbank}, {Strauss},
  {Yamada}, \& {Yamanoi}}]{Bosch2018}
{Bosch}, J., {Armstrong}, R., {Bickerton}, S., {et~al.} 2018, PASJ, 70, S5,
  \dodoi{10.1093/pasj/psx080}

\bibitem[{{Bowell} {et~al.}(1989){Bowell}, {Hapke}, {Domingue}, {Lumme},
  {Peltoniemi}, \& {Harris}}]{Bowell1989}
{Bowell}, E., {Hapke}, B., {Domingue}, D., {et~al.} 1989, in Asteroids II, ed.
  R.~P. {Binzel}, T.~{Gehrels}, \& M.~S. {Matthews}, 524--556

\bibitem[{{Buie} {et~al.}(2015){Buie}, {Olkin}, {Merline}, {Walsh}, {Levison},
  {Timerson}, {Herald}, {Owen}, {Abramson}, {Abramson}, {Breit}, {Caton},
  {Conard}, {Croom}, {Dunford}, {Dunford}, {Dunham}, {Ellington}, {Liu},
  {Maley}, {Olsen}, {Preston}, {Royer}, {Scheck}, {Sherrod}, {Sherrod},
  {Swift}, {Taylor}, \& {Venable}}]{Buie2015AJ....149..113B}
{Buie}, M.~W., {Olkin}, C.~B., {Merline}, W.~J., {et~al.} 2015, \aj, 149, 113,
  \dodoi{10.1088/0004-6256/149/3/113}

\bibitem[{{Chambers} {et~al.}(2017)}]{Chambers2017}
{Chambers}, K.~C., {et~al.} 2017, VizieR Online Data Catalog, II/349

\bibitem[{{Chang} {et~al.}(2016){Chang}, {Lin}, {Ip}, {Prince}, {Kulkarni},
  {Levitan}, {Laher}, \& {Surace}}]{Chang2016}
{Chang}, C.-K., {Lin}, H.-W., {Ip}, W.-H., {et~al.} 2016, \apjs, 227, 20,
  \dodoi{10.3847/0067-0049/227/2/20}

\bibitem[{{Chang} {et~al.}(2015){Chang}, {Ip}, {Lin}, {Cheng}, {Ngeow}, {Yang},
  {Waszczak}, {Kulkarni}, {Levitan}, {Sesar}, {Laher}, {Surace}, \&
  {Prince}}]{Chang2015}
{Chang}, C.-K., {Ip}, W.-H., {Lin}, H.-W., {et~al.} 2015, \apjs, 219, 27,
  \dodoi{10.1088/0067-0049/219/2/27}

\bibitem[{{Chang} {et~al.}(2019){Chang}, {Lin}, {Ip}, {Chen}, {Yeh},
  {Chambers}, {Magnier}, {Huber}, {Flewelling}, {Waters}, {Wainscoat}, \&
  {Schultz}}]{Chang2019}
{Chang}, C.-K., {Lin}, H.-W., {Ip}, W.-H., {et~al.} 2019, \apjs, 241, 6,
  \dodoi{10.3847/1538-4365/ab01fe}

\bibitem[{{Chapman}(1978)}]{Chapman1978NASCP2053..145C}
{Chapman}, C.~R. 1978, in NASA Conference Publication, Vol. 2053, NASA
  Conference Publication, ed. D.~{Morrison} \& W.~C. {Wells}, 145--160

\bibitem[{{Davis} {et~al.}(1985){Davis}, {Chapman}, {Weidenschilling}, \&
  {Greenberg}}]{Davis1985Icar...62...30D}
{Davis}, D.~R., {Chapman}, C.~R., {Weidenschilling}, S.~J., \& {Greenberg}, R.
  1985, \icarus, 62, 30, \dodoi{10.1016/0019-1035(85)90170-8}

\bibitem[{{DeMeo} \& {Carry}(2013)}]{Demeo2013Icar..226..723D}
{DeMeo}, F.~E., \& {Carry}, B. 2013, \icarus, 226, 723,
  \dodoi{10.1016/j.icarus.2013.06.027}

\bibitem[{Duda \& Hart(1972)}]{Duda1972}
Duda, R.~O., \& Hart, P.~E. 1972, Commun. ACM, 15, 11,
  \dodoi{10.1145/361237.361242}

\bibitem[{{Emery} {et~al.}(2011){Emery}, {Burr}, \& {Cruikshank}}]{Emery2011}
{Emery}, J.~P., {Burr}, D.~M., \& {Cruikshank}, D.~P. 2011, \aj, 141, 25,
  \dodoi{10.1088/0004-6256/141/1/25}

\bibitem[{{Fernandez} \& {Ip}(1984)}]{Fernandez1984}
{Fernandez}, J.~A., \& {Ip}, W.~H. 1984, \icarus, 58, 109,
  \dodoi{10.1016/0019-1035(84)90101-5}

\bibitem[{{Fleming} \& {Hamilton}(2000)}]{Fleming2000}
{Fleming}, H.~J., \& {Hamilton}, D.~P. 2000, \icarus, 148, 479,
  \dodoi{10.1006/icar.2000.6523}

\bibitem[{{Fraser} {et~al.}(2016{\natexlab{a}}){Fraser}, {Alexandersen},
  {Schwamb}, {Marsset}, {Pike}, {Kavelaars}, {Bannister}, {Benecchi}, \&
  {Delsanti}}]{Fraser2016}
{Fraser}, W., {Alexandersen}, M., {Schwamb}, M.~E., {et~al.}
  2016{\natexlab{a}}, \aj, 151, 158, \dodoi{10.3847/0004-6256/151/6/158}

\bibitem[{{Fraser} \& {Brown}(2018)}]{Fraser2018AJ....156...23F}
{Fraser}, W.~C., \& {Brown}, M.~E. 2018, \aj, 156, 23,
  \dodoi{10.3847/1538-3881/aac213}

\bibitem[{{Fraser} {et~al.}(2016{\natexlab{b}}){Fraser}, {Alexandersen},
  {Schwamb}, {Marsset}, {Pike}, {Kavelaars}, {Bannister}, {Benecchi}, \&
  {Delsanti}}]{trippy}
{Fraser}, W.~C., {Alexandersen}, M., {Schwamb}, M.~E., {et~al.}
  2016{\natexlab{b}}, {TRIPPy: Python-based Trailed Source Photometry}.
\newblock \doeprint{1605.010}

\bibitem[{{Furusawa} {et~al.}(2018){Furusawa}, {Koike}, {Takata}, {Okura},
  {Miyatake}, {Lupton}, {Bickerton}, {Price}, {Bosch}, {Yasuda}, {Mineo},
  {Yamada}, {Miyazaki}, {Nakata}, {Koshida}, {Komiyama}, {Utsumi},
  {Kawanomoto}, {Jeschke}, {Noumaru}, {Schubert}, {Iwata}, {Finet},
  {Fujiyoshi}, {Tajitsu}, {Terai}, \& {Lee}}]{Furusawa2018}
{Furusawa}, H., {Koike}, M., {Takata}, T., {et~al.} 2018, \pasj, 70, S3,
  \dodoi{10.1093/pasj/psx079}

\bibitem[{{Grundy} {et~al.}(2019){Grundy}, {Noll}, {Buie}, {Benecchi},
  {Ragozzine}, \& {Roe}}]{Grundy2019Icar..334...30G}
{Grundy}, W.~M., {Noll}, K.~S., {Buie}, M.~W., {et~al.} 2019, \icarus, 334, 30,
  \dodoi{10.1016/j.icarus.2018.12.037}

\bibitem[{{Harris}(1996)}]{Harris1996}
{Harris}, A.~W. 1996, in Lunar and Planetary Science Conference, Vol.~27, Lunar
  and Planetary Science Conference, 493

\bibitem[{{Harris} {et~al.}(1989){Harris}, {Young}, {Bowell}, {Martin},
  {Millis}, {Poutanen}, {Scaltriti}, {Zappala}, {Schober}, {Debehogne}, \&
  {Zeigler}}]{Harris1989}
{Harris}, A.~W., {Young}, J.~W., {Bowell}, E., {et~al.} 1989, Icarus, 77, 171,
  \dodoi{10.1016/0019-1035(89)90015-8}

\bibitem[{{Hirabayashi}(2015)}]{Hirabayashi2015}
{Hirabayashi}, M. 2015, \mnras, 454, 2249, \dodoi{10.1093/mnras/stv2017}

\bibitem[{{Holsapple}(2007)}]{Holsapple2007}
{Holsapple}, K.~A. 2007, \icarus, 187, 500,
  \dodoi{10.1016/j.icarus.2006.08.012}

\bibitem[{Hough(1959)}]{Hough1959}
Hough, P. V.~C. 1959, Conf. Proc. C, 590914, 554

\bibitem[{{Hu} {et~al.}(2021){Hu}, {Richardson}, {Zhang}, \& {Ji}}]{Hu2021}
{Hu}, S., {Richardson}, D.~C., {Zhang}, Y., \& {Ji}, J. 2021, \mnras,
  \dodoi{10.1093/mnras/stab412}

\bibitem[{{Kalup} {et~al.}(2021){Kalup}, {Moln{\'a}r}, {Kiss}, {Szab{\'o}},
  {P{\'a}l}, {Szak{\'a}ts}, {S{\'a}rneczky}, {Vink{\'o}}, {Szab{\'o}},
  {Kecskem{\'e}thy}, \& {Kiss}}]{Kalup2021ApJS..254....7K}
{Kalup}, C.~E., {Moln{\'a}r}, L., {Kiss}, C., {et~al.} 2021, \apjs, 254, 7,
  \dodoi{10.3847/1538-4365/abe76a}

\bibitem[{{Kawanomoto} {et~al.}(2018){Kawanomoto}, {Uraguchi}, {Komiyama},
  {Miyazaki}, {Furusawa}, {Finet}, {Hattori}, {Wang}, {Yasuda}, \&
  {Suzuki}}]{Kawanomoto2018}
{Kawanomoto}, S., {Uraguchi}, F., {Komiyama}, Y., {et~al.} 2018, \pasj, 70, 66,
  \dodoi{10.1093/pasj/psy056}

\bibitem[{{Komiyama} {et~al.}(2018){Komiyama}, {Obuchi}, {Nakaya}, {Kamata},
  {Kawanomoto}, {Utsumi}, {Miyazaki}, {Uraguchi}, {Furusawa}, {Morokuma},
  {Uchida}, {Miyatake}, {Mineo}, {Fujimori}, {Aihara}, {Karoji}, {Gunn}, \&
  {Wang}}]{Komiyama2018}
{Komiyama}, Y., {Obuchi}, Y., {Nakaya}, H., {et~al.} 2018, \pasj, 70, S2,
  \dodoi{10.1093/pasj/psx069}

\bibitem[{{Lykawka} \& {Horner}(2010)}]{Lykawka2010}
{Lykawka}, P.~S., \& {Horner}, J. 2010, \mnras, 405, 1375,
  \dodoi{10.1111/j.1365-2966.2010.16538.x}

\bibitem[{{Malhotra}(1995)}]{Malhotra1995}
{Malhotra}, R. 1995, \aj, 110, 420, \dodoi{10.1086/117532}

\bibitem[{{Mann} {et~al.}(2007){Mann}, {Jewitt}, \& {Lacerda}}]{Mann2007}
{Mann}, R.~K., {Jewitt}, D., \& {Lacerda}, P. 2007, \aj, 134, 1133,
  \dodoi{10.1086/520328}

\bibitem[{{Marchis} {et~al.}(2006){Marchis}, {Hestroffer}, {Descamps},
  {Berthier}, {Bouchez}, {Campbell}, {Chin}, {van Dam}, {Hartman}, {Johansson},
  {Lafon}, {Le Mignant}, {de Pater}, {Stomski}, {Summers}, {Vachier},
  {Wizinovich}, \& {Wong}}]{Marchis2006Natur.439..565M}
{Marchis}, F., {Hestroffer}, D., {Descamps}, P., {et~al.} 2006, \nat, 439, 565,
  \dodoi{10.1038/nature04350}

\bibitem[{{Marzari} \& {Scholl}(1998)}]{Marzari1998}
{Marzari}, F., \& {Scholl}, H. 1998, \icarus, 131, 41,
  \dodoi{10.1006/icar.1997.5841}

\bibitem[{{Masiero} {et~al.}(2009){Masiero}, {Jedicke}, {{\v{D}}urech}, {Gwyn},
  {Denneau}, \& {Larsen}}]{Masiero2009}
{Masiero}, J., {Jedicke}, R., {{\v{D}}urech}, J., {et~al.} 2009, \icarus, 204,
  145, \dodoi{10.1016/j.icarus.2009.06.012}

\bibitem[{{Miyazaki} {et~al.}(2018){Miyazaki}, {Komiyama}, {Kawanomoto}, {Doi},
  {Furusawa}, {Hamana}, {Hayashi}, {Ikeda}, {Kamata}, {Karoji}, {Koike},
  {Kurakami}, {Miyama}, {Morokuma}, {Nakata}, {Namikawa}, {Nakaya}, {Nariai},
  {Obuchi}, {Oishi}, {Okada}, {Okura}, {Tait}, {Takata}, {Tanaka}, {Tanaka},
  {Terai}, {Tomono}, {Uraguchi}, {Usuda}, {Utsumi}, {Yamada}, {Yamanoi},
  {Aihara}, {Fujimori}, {Mineo}, {Miyatake}, {Oguri}, {Uchida}, {Tanaka},
  {Yasuda}, {Takada}, {Murayama}, {Nishizawa}, {Sugiyama}, {Chiba}, {Futamase},
  {Wang}, {Chen}, {Ho}, {Liaw}, {Chiu}, {Ho}, {Lai}, {Lee}, {Jeng}, {Iwamura},
  {Armstrong}, {Bickerton}, {Bosch}, {Gunn}, {Lupton}, {Loomis}, {Price},
  {Smith}, {Strauss}, {Turner}, {Suzuki}, {Miyazaki}, {Muramatsu}, {Yamamoto},
  {Endo}, {Ezaki}, {Ito}, {Kawaguchi}, {Sofuku}, {Taniike}, {Akutsu}, {Dojo},
  {Kasumi}, {Matsuda}, {Imoto}, {Miwa}, {Suzuki}, {Takeshi}, \&
  {Yokota}}]{Miyazaki2018}
{Miyazaki}, S., {Komiyama}, Y., {Kawanomoto}, S., {et~al.} 2018, \pasj, 70, S1,
  \dodoi{10.1093/pasj/psx063}

\bibitem[{{Morbidelli} {et~al.}(2005){Morbidelli}, {Levison}, {Tsiganis}, \&
  {Gomes}}]{Morbidelli2005}
{Morbidelli}, A., {Levison}, H.~F., {Tsiganis}, K., \& {Gomes}, R. 2005, \nat,
  435, 462, \dodoi{10.1038/nature03540}

\bibitem[{{Mueller} {et~al.}(2010){Mueller}, {Marchis}, {Emery}, {Harris},
  {Mottola}, {Hestroffer}, {Berthier}, \& {di Martino}}]{Mueller2010}
{Mueller}, M., {Marchis}, F., {Emery}, J.~P., {et~al.} 2010, \icarus, 205, 505,
  \dodoi{10.1016/j.icarus.2009.07.043}

\bibitem[{{Nesvorn{\'y}} {et~al.}(2020){Nesvorn{\'y}}, {Vokrouhlick{\'y}},
  {Bottke}, {Levison}, \& {Grundy}}]{Nesvorny2020}
{Nesvorn{\'y}}, D., {Vokrouhlick{\'y}}, D., {Bottke}, W.~F., {Levison}, H.~F.,
  \& {Grundy}, W.~M. 2020, \apjl, 893, L16, \dodoi{10.3847/2041-8213/ab8311}

\bibitem[{{Nesvorn{\'y}} {et~al.}(2013){Nesvorn{\'y}}, {Vokrouhlick{\'y}}, \&
  {Morbidelli}}]{Nesvorny2013}
{Nesvorn{\'y}}, D., {Vokrouhlick{\'y}}, D., \& {Morbidelli}, A. 2013, \apj,
  768, 45, \dodoi{10.1088/0004-637X/768/1/45}

\bibitem[{{Polishook} {et~al.}(2012){Polishook}, {Ofek}, {Waszczak},
  {Kulkarni}, {Gal-Yam}, {Aharonson}, {Laher}, {Surace}, {Klein}, {Bloom},
  {Brosch}, {Prialnik}, {Grillmair}, {Cenko}, {Kasliwal}, {Law}, {Levitan},
  {Nugent}, {Poznanski}, \& {Quimby}}]{Polishook2012MNRAS.421.2094P}
{Polishook}, D., {Ofek}, E.~O., {Waszczak}, A., {et~al.} 2012, \mnras, 421,
  2094, \dodoi{10.1111/j.1365-2966.2012.20462.x}

\bibitem[{{Romanishin} \& {Tegler}(2018)}]{Romanishin2018}
{Romanishin}, W., \& {Tegler}, S.~C. 2018, \aj, 156, 19,
  \dodoi{10.3847/1538-3881/aac210}

\bibitem[{{Rozitis} \& {Green}(2013)}]{Rozitis2013MNRAS.430.1376R}
{Rozitis}, B., \& {Green}, S.~F. 2013, \mnras, 430, 1376,
  \dodoi{10.1093/mnras/sts723}

\bibitem[{{Rubincam}(2000)}]{Rubincam2000}
{Rubincam}, D.~P. 2000, \icarus, 148, 2, \dodoi{10.1006/icar.2000.6485}

\bibitem[{{Ryan} {et~al.}(2017){Ryan}, {Sharkey}, \& {Woodward}}]{Ryan2017}
{Ryan}, E.~L., {Sharkey}, B. N.~L., \& {Woodward}, C.~E. 2017, \aj, 153, 116,
  \dodoi{10.3847/1538-3881/153/3/116}

\bibitem[{{Sonnett} {et~al.}(2015){Sonnett}, {Mainzer}, {Grav}, {Masiero}, \&
  {Bauer}}]{Sonnett2015}
{Sonnett}, S., {Mainzer}, A., {Grav}, T., {Masiero}, J., \& {Bauer}, J. 2015,
  \apj, 799, 191, \dodoi{10.1088/0004-637X/799/2/191}

\bibitem[{{Szab{\'o}} {et~al.}(2017){Szab{\'o}}, {P{\'a}l}, {Kiss}, {Kiss},
  {Moln{\'a}r}, {Hanyecz}, {Plachy}, {S{\'a}rneczky}, \&
  {Szab{\'o}}}]{Szabo2017}
{Szab{\'o}}, G.~M., {P{\'a}l}, A., {Kiss}, C., {et~al.} 2017, \aap, 599, A44,
  \dodoi{10.1051/0004-6361/201629401}

\bibitem[{{Szab{\'o}} {et~al.}(2020){Szab{\'o}}, {Kiss}, {Szak{\'a}ts},
  {P{\'a}l}, {Moln{\'a}r}, {S{\'a}rneczky}, {Vink{\'o}}, {Szab{\'o}}, {Marton},
  \& {Kiss}}]{Szabo2020}
{Szab{\'o}}, G.~M., {Kiss}, C., {Szak{\'a}ts}, R., {et~al.} 2020, \apjs, 247,
  34, \dodoi{10.3847/1538-4365/ab6b23}

\bibitem[{{Warner} {et~al.}(2009){Warner}, {Harris}, \& {Pravec}}]{Warner2009}
{Warner}, B.~D., {Harris}, A.~W., \& {Pravec}, P. 2009, \icarus, 202, 134,
  \dodoi{10.1016/j.icarus.2009.02.003}

\bibitem[{{Weissman}(1986)}]{Weissman1986Natur.320..242W}
{Weissman}, P.~R. 1986, \nat, 320, 242, \dodoi{10.1038/320242a0}

\bibitem[{{Willmer}(2018)}]{Willmer2018}
{Willmer}, C. N.~A. 2018, \apjs, 236, 47, \dodoi{10.3847/1538-4365/aabfdf}

\bibitem[{{Wong} \& {Brown}(2015)}]{Wong2015}
{Wong}, I., \& {Brown}, M.~E. 2015, \aj, 150, 174,
  \dodoi{10.1088/0004-6256/150/6/174}

\bibitem[{{Wong} {et~al.}(2014){Wong}, {Brown}, \& {Emery}}]{Wong2014}
{Wong}, I., {Brown}, M.~E., \& {Emery}, J.~P. 2014, \aj, 148, 112,
  \dodoi{10.1088/0004-6256/148/6/112}

\bibitem[{{Yoshida} \& {Terai}(2017)}]{Yoshida2017}
{Yoshida}, F., \& {Terai}, T. 2017, \aj, 154, 71,
  \dodoi{10.3847/1538-3881/aa7d03}

\end{thebibliography}

\end{document}